\documentclass[11pt]{amsart}

\usepackage[margin=1in, includefoot, footskip=30pt]{geometry}
\usepackage{amsmath, amsfonts, amssymb, mathrsfs, enumitem}
\usepackage{setspace, tikz, pgfplots}
\usepackage{stackrel, nicefrac, epsfig}
\usepackage{algorithm}
\usepackage[noend]{algpseudocode}
\usepackage[hyperfootnotes=false]{hyperref}
\usepackage[labelfont=rm]{subfig}
\pgfplotsset{compat=1.13}
\newtheorem{rmk}{Remark}[section]
\DeclareMathOperator*{\argmax}{argmax}
\DeclareMathOperator*{\tr}{tr}
\newcommand{\eff}{{\mathrm{eff}}}
\usepackage{caption}
\title[Reinforcement Learning Using Quantum Boltzmann Machines]
{Reinforcement Learning\\Using Quantum Boltzmann Machines}
\date{\today}

\keywords{Reinforcement learning, Machine learning, Neuro-dynamic programming,
Markov decision process, Quantum Monte Carlo simulation, Simulated quantum
annealing, Restricted Boltzmann machine, Deep Boltzmann machine, General
Boltzmann machine, Quantum Boltzmann machine}

\author[D. Crawford]{Daniel Crawford}
\author[A. Levit]{Anna Levit}
\author[N.Ghadermarzy]{Navid Ghadermarzy}
\author[J. S. Oberoi]{Jaspreet S. Oberoi}
\author[P. Ronagh]{Pooya Ronagh}

\email[Daniel Crawford]{daniel.crawford@1qbit.com}
\email[Anna Levit]{anna.levit@1qbit.com}
\email[Navid Ghadermarzy]{navidgh@math.ubc.ca}
\email[Jaspreet S. Oberoi]{jaspreet.oberoi@1qbit.com}
\email[Pooya Ronagh]{pooya.ronagh@1qbit.com}

\address[Daniel Crawford, Anna Levit, Jaspreet S. Oberoi, Pooya Ronagh]{
1QB Information Technologies (1QBit)
\vspace*{-3mm}}
\address[Navid Ghadermarzy]{
Department of Mathematics, University of British Columbia
\vspace*{-3mm}}
\address[Jaspreet S. Oberoi]{
School of Engineering Science, Simon Fraser University
\vspace*{-3mm}}
\address[Pooya Ronagh]{
Institute for Quantum Computing and Department of Physics and Astronomy,
University of Waterloo}

\begin{document}

\begin{abstract}
We investigate whether quantum annealers with select chip layouts can outperform
classical computers in reinforcement learning tasks. We associate a transverse
field Ising spin Hamiltonian with a layout of qubits similar to that of a deep
Boltzmann machine (DBM) and use simulated quantum annealing (SQA) to
numerically simulate quantum sampling from this system. We design a
reinforcement learning algorithm in which the set of visible nodes representing
the states and actions of an optimal policy are the first and last layers of the
deep network. In absence of a transverse field, our simulations show that DBMs
are trained more effectively than restricted Boltzmann machines (RBM) with the
same number of nodes. We then develop a framework for training the network as a
quantum Boltzmann machine (QBM) in the presence of a significant transverse
field for reinforcement learning. This method also outperforms the reinforcement
learning method that uses RBMs.
\end{abstract}

\maketitle
\onehalfspacing
\let\thefootnote\relax\footnotetext{
\emph{Published in:} Quantum Information and Computation,
{Vol.~18}, {No.~1\&2}, {pp.~0051--0074}, Rinton~Press (2018).}

\section{Introduction}

Recent theoretical extensions of the quantum adiabatic theorem
\cite{PhysRevA.71.012331, Avron2012, 136726301412123016, 2016arXiv161201505B,
PhysRevA.93.032118} suggest the possibility of using quantum devices with
manufactured spins \cite{dwnature, martini} as samplers of the instantaneous
steady states of quantum systems. With this motivation, we consider
reinforcement learning as the computational task of interest, and design a
method of reinforcement learning consisting of sampling from a layout of
quantum bits similar to that of a deep Boltzmann machine (DBM) (see Fig.~\ref
{dbm-graph} for a graphical representation). We use simulated quantum annealing
(SQA) to demonstrate the advantage of reinforcement learning using deep
Boltzmann machines and quantum Boltzmann machines over their classical
counterpart, for small problem instances.

\def\layersep{3cm}
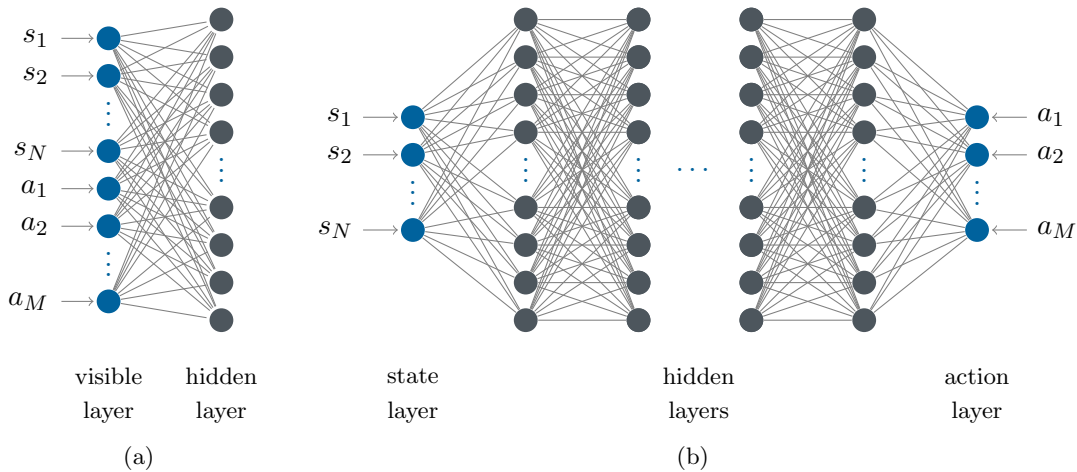
\begin{figure*}[t]
\subfloat[]{\label{rbm-graph}
\begin{tikzpicture}[scale=0.5, shorten >=1pt,-,draw=black!50, node
distance=\layersep]
    \tikzstyle{every pin edge}=[<-,shorten <=1pt]
    \tikzstyle{neuron}=[circle,fill=black!25,minimum size=9pt,inner sep=0pt]
    \tikzstyle{input neuron}=[neuron, fill={rgb:red,0;green,117;blue,188}];
    \tikzstyle{output neuron}=[neuron, fill=red!75];
    \tikzstyle{hidden neuron}=[neuron, fill={rgb:red,114;green,128;blue,138}];
    \tikzstyle{annot} = [text width=4em, text centered]
    \node[input neuron, pin=left:$s_1$] (I-1) at (0,-1) {};
    \node[input neuron, pin=left:$s_2$] (I-2) at (0,-2) {};
    \node[input neuron, pin=left:$s_N$] (I-4) at (0,-4) {};
    \node[input neuron, pin=left:$a_1$] (I-5) at (0,-5) {};
    \node[input neuron, pin=left:$a_2$] (I-6) at (0,-6) {};
    \node[input neuron, pin=left:$a_M$] (I-8) at (0,-8) {};
    \node[color={rgb:red,0;green,117;blue,188}, align=left] at (0,-2.8)
{$\vdots$};
    \node[color={rgb:red,0;green,117;blue,188}, align=left] at (0,-6.8)
{$\vdots$};
    \node[color={rgb:red,0;green,117;blue,188}, align=left] at (\layersep,-4.3)
{$\vdots$};
    \foreach \name / \y in {1,2,3,4,6,7,8,9}
        \path[yshift=0.5cm]
            node[hidden neuron] (H-\name) at (\layersep,-\y cm) {};
    \foreach \source in {1,2,4,5,6,8}
        \foreach \dest in {1,2,3,4,6,7,8,9}
            \path (I-\source) edge (H-\dest);
    \node[annot,below of=H-1, node distance=5cm] (hl){\footnotesize{hidden
layer}};
    \node[annot,left of=hl, node distance=1.5cm] {\footnotesize{visible layer}};
	\node[] at (0,-10.5){};
\end{tikzpicture}}
\def\layersep{1.5cm}
\subfloat[]{\label{dbm-graph}
\begin{tikzpicture}[scale=1,-,draw=black!50, node distance=\layersep]
\tikzstyle{every pin edge}=[<-,shorten <=1pt]
\tikzstyle{neuron}=[circle,fill=black!25,minimum size=9pt,inner sep=0pt]
\tikzstyle{input neuron}=[neuron, fill={rgb:red,0;green,117;blue,188}];
\tikzstyle{output neuron}=[neuron, fill=red!75];
\tikzstyle{hidden neuron}=[neuron, fill={rgb:red,114;green,128;blue,138}];
\tikzstyle{annot} = [text width=4em, text centered]
\node[input neuron, pin=left:$s_{1}$] (I-1) at (0,-1-0.3) {};
\node[input neuron, pin=left:$s_{2}$] (I-2) at (0,-1.5-0.3) {};
\node[input neuron, pin=left:$s_{N}$] (I-4) at (0,-2.5-0.3) {};
\node[color={rgb:red,0;green,117;blue,188}, align=left] at (0,-1.9-0.3)
{$\vdots$};
\node[color={rgb:red,0;green,117;blue,188}, align=left] at
(\layersep*5,-1.9-0.3) {$\vdots$};
\node[color={rgb:red,0;green,117;blue,188}, align=left] at (\layersep,-1.9)
{$\vdots$};
\node[color={rgb:red,0;green,117;blue,188}, align=left] at (\layersep*2,-1.9)
{$\vdots$};
\node[color={rgb:red,0;green,117;blue,188}, align=left] at (\layersep*3,-1.9)
{$\vdots$};
\node[color={rgb:red,0;green,117;blue,188}, align=left] at (\layersep*4,-1.9)
{$\vdots$};
\node[color={rgb:red,0;green,117;blue,188}, align=left] at (\layersep*2.5,-2.0)
{$\dots$};
\foreach \name / \y in {1,2,3,4,6,7,8,9}
\path[yshift=0.5cm]
node[hidden neuron] (H-\name) at (\layersep,-\y*0.5 cm) {};
\foreach \name / \y in {1,2,3,4,6,7,8,9}
\path[yshift=0.5cm]
node[hidden neuron,right of=H-3] (H1-\name) at (\layersep,-\y*0.5 cm) {};
\foreach \name / \y in {1,2,3,4,6,7,8,9}
\path[yshift=0.5cm]
node[hidden neuron,right of=H1-3] (H2-\name) at (\layersep,-\y*0.5 cm) {};
\foreach \name / \y in {1,2,3,4,6,7,8,9}
\path[yshift=0.5cm]
node[hidden neuron,right of=H2-3] (HN1-\name) at (\layersep*2,-\y*0.5 cm) {};
\foreach \name / \y in {1,2,3,4,6,7,8,9}
\path[yshift=0.5cm]
node[hidden neuron,right of=H2-3] (HN1-\name) at (\layersep*2,-\y*0.5 cm) {};
\foreach \name / \y in {1,2,3,4,6,7,8,9}
\path[yshift=0.5cm]
node[hidden neuron,right of=H2-3] (HN-\name) at (\layersep*3,-\y*0.5 cm) {};
\node[input neuron, pin=right:$a_{1}$] (V2-1) at (\layersep*5,-1-0.3) {};
\node[input neuron, pin=right:$a_{2}$] (V2-2) at (\layersep*5,-1.5-0.3) {};
\node[input neuron, pin=right:$a_{M}$] (V2-4) at (\layersep*5,-2.5-0.3) {};
\foreach \source in {1,2,4}
\foreach \dest in {1,2,3,4,6,7,8,9}
\path (I-\source) edge (H-\dest);
\foreach \source in {1,2,3,4,6,7,8,9}
\foreach \dest in {1,2,3,4,6,7,8,9}
\path (H-\source) edge (H2-\dest);
\foreach \source in {1,2,3,4,6,7,8,9}
\foreach \dest in {1,2,3,4,6,7,8,9}
\path (HN1-\source) edge (HN-\dest);
\foreach \source in {1,2,3,4,6,7,8,9}
\foreach \dest in {1,2,4}
\path (HN-\source) edge (V2-\dest);
\node[annot,above of=H2-1, node distance=0.5cm] (h1) {};
\node[annot,below of=h1, node distance=5.5cm] (labels){};
\node[annot,left of=labels, node distance=3cm] {\footnotesize{state \\layer}};
\node[annot,right of=labels, node distance=0.8cm] {\footnotesize{hidden
layers}};
\node[annot,right of=labels, node distance=4.5cm] {\footnotesize{action layer}};
\node[annot, above of=HN1-1, node distance=0.5cm](hk-11){};
\node[annot, above of=HN-1, node distance=0.5cm](hk1){};
\node[annot, below of=h1, node distance=5.5cm](h1N1){};
\node[annot, right of=h1N1, node distance=1.5cm](h2N2){};
\node[annot, right of=h2N2, node distance=1.5cm](hK-1NK-1){};
\node[annot, right of=hK-1NK-1, node distance=1.65cm](hKNK){};
\end{tikzpicture}}
\vspace*{13pt}
\captionof{figure}{(a) The general RBM layout used in RBM-based reinforcement
learning. The visible layer on the left consists of state and action nodes, and
is connected to the hidden layer, forming a complete bipartite graph. (b) The
general DBM layout used in DBM-based reinforcement learning. The visible nodes
on the left represent states and the visible nodes on the right represent
actions. The training procedure captures the correlations between states and
actions in the weights of the edges between the nodes.
\label{dbmgraph}}
\end{figure*}

Reinforcement learning (\cite{sutton-book}, known also as neuro-dynamic
programming \cite{bertsekas1996neuro}) is an area of optimal control theory at
the intersection of approximate dynamic programming and machine learning. It
has been used successfully for many applications, in fields such as engineering
\cite{derhami2013applying, syafiie2007model}, sociology
\cite{erev1998predicting, shteingart2014reinforcement}, and economics
\cite{matsui2011compound, sui2010reinforcement}.

It is important to differentiate between reinforcement learning and common
streams of research in machine learning. For instance, in supervised learning,
the learning is facilitated by training samples provided by a source external
to the agent and the computer. In reinforcement learning, the training samples
are provided only by the interaction of the agent itself with the environment.
For example, in a motion planning problem in an uncharted territory, it is
desired that the agent learns in the fastest possible way to navigate correctly, 
with the fewest blind decisions required to be made. This is known as
the dilemma of \emph{exploration versus exploitation}; that is, neither
exploration nor exploitation can be pursued exclusively without facing a
penalty or failing at the task. The goal is hence not only to design an
algorithm that eventually converges to an optimal policy, but for it to be able
to generate good policies early in the learning process. We refer the reader to
\cite[Ch.~1.1]{sutton-book} for a thorough introduction to use cases and
problem scenarios addressed by reinforcement learning.

The core idea in reinforcement learning is defining an operator on the Banach
space of real-valued functions on the set of states of a system such that a
fixed point of the operator carries information about an optimal policy of
actions for a finite or infinite number of decision epochs. A numerical method
for computing this fixed point is to explore this function space by travelling
in a direction that minimizes the distance between two consecutive applications
of the contraction mapping operator \cite{bertsekas1996neuro}.

This optimization task, called \emph{learning} in the context of reinforcement
learning, can be performed by locally parametrizing the above function space
using a set of auxiliary variables, and applying a gradient method to these
variables. One approach for such a parametrization, due to \cite{HintonRBM}, is
to use the weights of a restricted Boltzmann machine (RBM) (see
Fig.~\ref{rbm-graph}) as the parameters, and the free energy of the RBM as an
approximator for the elements in the function space. The descent direction is
then calculated in terms of the expected values of the nodes of the RBM.

It follows from the universal approximation theorem \cite{hornik1989multilayer}
that RBMs can approximate any joint distribution over binary variables
\cite{martens2013representational, le2008representational}. However, in the
context of reinforcement learning, RBMs are not necessarily the best choice for
approximating \mbox{Q-functions} relating to Markov decision processes because
RBMs may require an exponential number of hidden variables with respect to the
number of visible variables in order to approximate the desired joint
distribution \cite{martens2013representational, le2008representational}. On the
other hand, DBMs have the potential to model higher-order dependencies than
RBMs, and are more robust than deep belief networks
\cite{salakhutdinov2009deep}.

One may, therefore, consider replacing the RBM with other graphical models and
investigating the performance of the models in the learning process. Except in
the case of RBMs, calculating statistical data from the nodes of a graphical
model amounts to sampling from a Boltzmann distribution, creating a bottleneck
in the learning procedure. Therefore, any improvement in the efficiency of
Boltzmann distribution sampling is beneficial for reinforcement learning and
machine learning in general.

\def\boxsep{0.5}
\begin{figure*}[t!]
\centering
\subfloat[]{\label{fig:grid-det}
\begin{tikzpicture}[scale=0.7]
\draw[ultra thick] (0,0) rectangle (5,-3);
\foreach \row in {0,...,5}{
	\foreach \column in {0,...,3} {
		\draw[dotted] (0,0) rectangle +(1*\row, -1*\column);}}
\node[align=left] at (0.5,-0.5) {$R$};
\node[align=left] at (2.5,-1.5) {$W$}; 
\node[align=left] at (2.5, -2.5) {$P$};
\end{tikzpicture}}
\subfloat[]{\label{fig:grid-stoc}
\begin{tikzpicture}[scale=0.7]
\draw[ultra thick] (0,0) rectangle (5,-3);
\foreach \row in {0,...,5}{
	\foreach \column in {0,...,3} {
		\draw[dotted] (0,0) rectangle +(1*\row, -1*\column);}}
\node[align=left] at (0.5,-0.5) {$R$};
\node[align=left] at (2.5,-1.5) {$W$};
\node[align=left] at (0.5, -2.5) {$\mathcal{R}$};
\node[align=left] at (4.5, -0.5) {$\mathcal{R}$};
\node[align=left] at (2.5, -2.5) {$P$};
\end{tikzpicture}}
\subfloat[]{\label{fig:grid-sol}
\begin{tikzpicture}[scale=0.7]
\draw[ultra thick] (0,-3.5) rectangle (5,-6.5);
\foreach \row in {0,...,5}{
	\foreach \column in {0,...,3} {
		\draw[dotted] (0,-3.5) rectangle +(1*\row, -1*\column);}}
\node[align=left] at (0.5,-4) {$\circlearrowleft$};
\node[align=left] at (2.5,-5) {$W$};
\node[align=left] at (0.5, -6) {$\uparrow$};
\node[align=left] at (4.5, -4) {$\leftarrow$};
\node[align=left] at (2.5, -6) {$\leftarrow$};
\node[align=left] at (0.5,-5) {$\uparrow$};
\node[align=left] at (1.5,-4) {$\leftarrow$};
\node[align=left] at (1.4,-5) {$\leftarrow$};
\node[align=left] at (1.5,-4.9) {$\uparrow$};
\node[align=left] at (1.4,-6) {$\leftarrow$};
\node[align=left] at (1.5,-5.9) {$\uparrow$};
\node[align=left] at (2.5,-4) {$\leftarrow$};
\node[align=left] at (3.5,-4) {$\leftarrow$};
\node[align=left] at (3.5,-5) {$\uparrow$};
\node[align=left] at (3.5,-6) {$\uparrow$};
\node[align=left] at (4.4,-5) {$\leftarrow$};
\node[align=left] at (4.5,-4.9) {$\uparrow$};
\node[align=left] at (4.4,-6) {$\leftarrow$};
\node[align=left] at (4.5,-5.9) {$\uparrow$};
\end{tikzpicture}}\\
\subfloat[]{
\hspace{0.1in}
\begin{tikzpicture}[scale=0.7] \label{fig:wind-ker}
\draw[densely dotted,line width = .3mm] (1, 0+\boxsep) rectangle
(2,-1+\boxsep);  
\draw[densely dotted,line width = .3mm] (1,-1) rectangle (2,-2); 
\draw[densely dotted,line width = .3mm] (1, -2-\boxsep) rectangle
(2,-3-\boxsep); 
\draw[densely dotted,line width = .3mm] (0-\boxsep, -1) rectangle
(1-\boxsep,-2); 
\draw[densely dotted,line width = .3mm] (2+\boxsep, -1) rectangle
(3+\boxsep,-2); 
\draw [-to,shorten >=-1pt,thick] (1.5, -1) -- (1.5, 0);  
\draw [-to,shorten >=-1pt,thick] (1.5, -2) -- (1.5, -3);  
\draw [-to,shorten >=-1pt,thick] (1, -1.5) -- (0, -1.5);  
\draw [-to,shorten >=-1pt,thick] (2, -1.5) -- (3, -1.5);  
\node[align=left] at (1.5, -1.5) {$\rightarrow$};
\node[align=left] at (1.5, 0.25) {\scriptsize $0.05$}; 
\node[align=left] at (1.5, -3.25) {\scriptsize $0.05$}; 
\node[align=left] at (3, -1.2) {\scriptsize $0.8$}; 
\node[align=left] at (0, -1.2) {\scriptsize $0.05$}; 
\node[align=left] at (1.5, -1.2) {\scriptsize $0.05$}; 
\def\horsep{5}                                    
\draw[densely dotted,line width = .3mm] (1+\horsep, 0+\boxsep) rectangle
(2+\horsep,-1+\boxsep);  
\draw[densely dotted,line width = .3mm] (1+\horsep,-1) rectangle
(2+\horsep,-2); 
\draw[densely dotted,line width = .3mm] (1+\horsep, -2-\boxsep) rectangle
(2+\horsep,-3-\boxsep); 
\draw[densely dotted,line width = .3mm] (0-\boxsep+\horsep, -1) rectangle
(1-\boxsep+\horsep,-2); 
\draw[densely dotted,line width = .3mm] (2+\boxsep+\horsep, -1) rectangle
(3+\boxsep+\horsep,-2); 
\draw [-to,shorten >=-1pt,thick] (1.5+\horsep, -1) -- (1.5+\horsep, 0);  
\draw [-to,shorten >=-1pt,thick] (1+\horsep, -1.5) -- (0+\horsep, -1.5);  
\draw [-to,shorten >=-1pt,thick] (2+\horsep, -1.5) -- (3+\horsep, -1.5);  %
right
\node[align=left] at (1.5+\horsep, -1.5) {$\rightarrow$};
\node[align=left] at (1.5+\horsep, -3) {$W$};
\node[align=left] at (1.5+\horsep, 0.25) {\scriptsize $\nicefrac{1}{15}$}; 
\node[align=left] at (1.5+\horsep, -1.2) {\scriptsize $\nicefrac{1}{15}$}; 
\node[align=left] at (3+\horsep, -1.2) {\scriptsize $0.8$}; 
\node[align=left] at (0+\horsep, -1.2) {\scriptsize $\nicefrac{1}{15}$}; 
\end{tikzpicture}}
\vspace*{13pt}
\captionof{figure}{(a) A $3 \times 5$ maze. $W$ represents a wall, $R$ is a
positive real number representing a reward, and $P$ is a real number
representing a penalty. (b) The previous maze with two additional stochastic
rewards. (c) The set of all optimal actions for each cell of the maze in Fig
(a). An optimal traversal policy is a choice of any combination of these
actions. (d) A sample conditional state transition probability for a windy
problem with no obstacles (left), and with a wall present (right).}
\end{figure*}
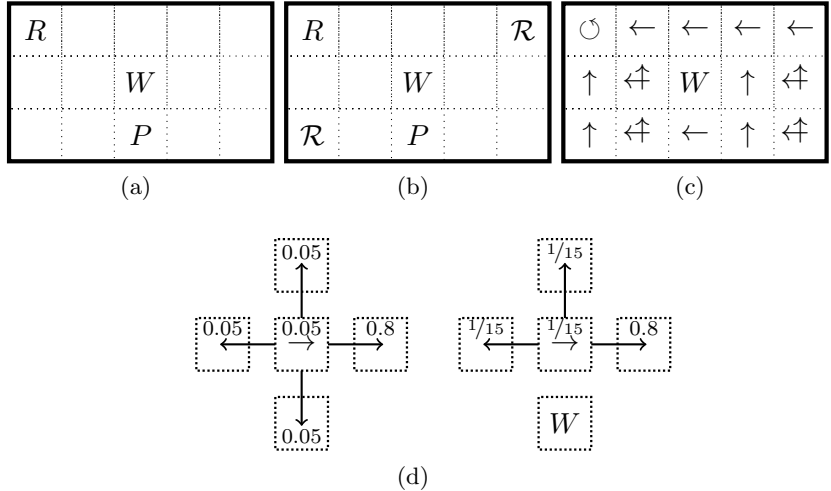

As we explain in what follows, DBMs are good candidates for reinforcement
learning tasks. Moreover, an important advantage of a DBM layout for a quantum
annealing system is that the proximity and couplings of the qubits in the layout
are similar to those of a sequence of bipartite blocks in D-Wave Systems'
devices \cite{PhysRevB82024511}, and it is therefore feasible that such layouts
could be manufactured in the near future. In addition, embedding Boltzmann
machines in larger quantum annealer architectures is problematic when
excessively large weights and biases are needed to emulate logical nodes of the
Boltzmann machine using chains and clusters of physical qubits. These are the
reasons why, instead of attempting to embed a Boltzmann machine structure on an
existing quantum annealing system as in \cite{1609.02542, 2015arXiv151006356A,
Denil:2011, 2016PhRvA94b2308B}, we work under the assumption that the network
itself is the native connectivity graph of a near-future quantum annealer, and,
using numerical simulations, we attempt to understand its applicability to
reinforcement learning.

We also refer the reader to current trends in machine learning using quantum
circuits, specifically, \cite{PhysRevLett.117.130501} and \cite{4579244} for
reinforcement learning, and \cite{MR3559656} and \cite{2016arXiv161205204K} for
training quantum Boltzmann machines with applications in deep learning and
tomography. To the best of our knowledge, the present paper complements the
literature on quantum machine learning as the first proposal on reinforcement
learning using adiabatic quantum computation.

Quantum Monte Carlo (QMC) numerical simulations have been found to be useful in
simulating time-dependant quantum systems. Simulated quantum annealing (SQA)
\cite{Crosson, 1411.5693}, one of the many flavours of QMC methods, is based on
the Suzuki--Trotter expansion of the path integral representation of the
Hamiltonian of Ising spin models in the presence of a transverse field driver
Hamiltonian. Even though the efficiency of SQA for finding the ground state of
an Ising model is topologically obstructed \cite{1302.5733}, we consider the
samples generated by SQA to be good approximations of the Boltzmann distribution
of the quantum Hamiltonian \cite{morita2006convergence}. Experimental studies
have shown similarities in the behaviour of SQA and that of quantum annealing
\cite{1510.08057, 2014arXiv1409.3827A} and its physical realization by D-Wave Systems \cite{arXiv:1509.02562v2, arXiv:1401.7087v2}.

We expect that when SQA is set such that the final strength of the transverse
field is negligible, the distribution of the samples approaches the classical
limit one expects to observe in absence of the transverse field. Another
classical algorithm which can be used to obtain samples from the Boltzmann
distribution is conventional simulated annealing (SA), which is based on thermal
annealing. Note that this algorithm can be used to create Boltzmann
distributions from the Ising spin model only in the absence of a transverse
field. It should, therefore, be possible to use SA or SQA to approximate the
Boltzmann distribution of a classical Boltzmann machine. However, unlike in the
case of SA, it is possible to use SQA not only to approximate the Boltzmann
distribution of a classical Boltzmann machine, but also that of a graphical
model in which the energy operator is a quantum Hamiltonian in the presence of a
transverse field. These graphical models, called quantum Boltzmann machines
(QBM), were first introduced in \cite{1601.02036}.

We use SQA simulations to provide evidence that a quantum annealing device that
approximates the distribution of a DBM or a QBM may improve the learning process
compared to a reinforcement learning method that uses classical RBM techniques.
Other studies have shown that SQA is more efficient than thermal SA
\cite{Crosson, 1411.5693}. Therefore, our method, used in conjunction with SQA,
can also be viewed as a quantum-inspired approach for reinforcement learning.

What distinguishes our work from current trends in quantum machine learning is
that (i) we consider the use of quantum annealing in reinforcement learning
applications rather than frequently studied classification or recognition
problems; (ii) using SQA-based numerical simulations, we assume that the
connectivity graph of a DBM directly maps to the native layout of a feasible
quantum annealer; and (iii) the results of our experiments using SQA to simulate
the sampling of an entangled system of spins suggest that using quantum
annealers in reinforcement learning tasks can offer an advantage over thermal
sampling.

\section{Preliminaries} \label{sec:preliminaries}

\subsection{Adiabatic Evolution of Open Quantum Systems}
\label{sec:open}

The evolution of a quantum system under a slowly changing time-dependent
Hamiltonian is characterized by the quantum adiabatic theorem (QAT). QAT has a
long history going back to the work of Born and Fock~\cite{1928ZPhy}.
Colloquially, QAT states that a system remains close to its instantaneous
steady state provided there is a gap between the eigenenergy of the steady
state and the rest of the Hamiltonian's spectrum at every point in time if the
evolution is sufficiently slow. This result motivated \cite{PhysRevE.58.5355}
and \cite{2000quant.ph1106F} to introduce the closely related paradigms of
quantum computing known as quantum annealing (QA) and adiabatic quantum
computation (AQC).

QA and AQC, in turn, inspired efforts in the manufacturing of physical
realizations of adiabatic evolution via quantum hardware (\cite{dwnature}). In
reality, the manufactured chips operate at nonzero temperature and are not
isolated from their environment. Therefore, the existing adiabatic theory did
not describe the behaviour of these machines. A contemporary investigation in
quantum adiabatic theory was thus initiated to study adiabaticity in open
quantum systems (\cite{PhysRevA.71.012331, Avron2012, 136726301412123016,
2016arXiv161201505B, PhysRevA.93.032118}). These references prove adiabatic
theorems to various degrees of generality and under a variety of assumptions 
about the system.

In fact, \cite{Avron2012} develops an adiabatic theory for equations of the form
\begin{equation}\label{eq:avron-master}
\varepsilon \dot x (s) = L(s) x(s),
\end{equation}
where $L$ is a family of linear operators on a Banach space and $L(s)$ is a
generator of a contraction semigroup for every $s$. This provides a general
framework that encompasses many adiabatic theorems, including that of classical
stochastic systems, all the way to quantum evolutions of open systems generated
by Lindbladians. The manifold of instantaneous stationary states is identical to
$\ker (L(s))$, and \cite{Avron2012} shows that the dynamics of the system are
parallel-transported along this manifold as $\varepsilon \to 0$.

An example of \eqref{eq:avron-master} is the case in which the Banach space is
the space of bounded operators on a Hilbert space, and in this case we study the
evolution of the density matrix $\rho$ of a quantum system. The Lindbladian is
defined via the adjoint action of a Hermitian $H$ on $\rho$, and couplings to
the heat bath are represented via a family of operators $\Gamma_\alpha$ with
$\sum_\alpha \Gamma_\alpha^* \Gamma_\alpha$ being bounded:
$$L \mathcal \rho = -i [H , \rho] + \frac{1}2 \sum_{\alpha} ([\Gamma_\alpha
\rho, \Gamma_\alpha^\ast] + [\Gamma_\alpha, \rho \Gamma_\alpha^\ast]).$$
In the work of \cite{Avron2012}, it was then proven that $\rho(s)$ is
parallel-transported along $\ker (L(s))$, and that if $L$:
\begin{enumerate}[label=(\roman*), leftmargin=2pc, itemsep=0pt]
\item is the generator of a contraction semigroup;
\item has closed and complementary range and kernel;
\item is $C^k$ with respect to $s$; and
\item is constant near the endpoints $s= 0$ and $s= 1$;
\end{enumerate}
then the solution to \eqref{eq:avron-master} with initial condition in $\ker
(L(0))$ deviates only in $O(\varepsilon^k)$ from $\ker(L(1))$ at $s= 1$.

The authors of \cite{PhysRevA.93.032118} focuses on estimating the adiabatic
error in terms of the physical parameters of the theory. In particular, they
study the case of a quantum system coupled to a thermal bath satisfying the 
Kubo--Martin--Schwinger (KMS) condition. Given a distance $\delta$, in order
for the norm of the solution of \eqref{eq:avron-master} to stay $\delta$-close
to the instantaneous steady state of the system at $s=1$, they show that
$\varepsilon$ has to decrease at a rate of $O(\lambda^2)$, where $\lambda$
denotes the smallest nonzero eigenvalue in $L$. Note that the KMS condition
implies that the Gibbs state $\exp (-\beta H(s)) / \tr [\exp (-\beta H(s))]$ is,
in fact, in $\ker (L(s))$.

This stream of research suggests promising opportunities to use quantum
annealers to sample from the Gibbs state of a quantum Hamiltonian using
adiabatic evolution. In this paper, the transverse field Ising model (TFIM) has
been the centre of attention. In practice, due to additional complications in
quantum annealing (e.g., level crossings and gap closure), the samples gathered
from the quantum annealer are far from the Gibbs state of the final Hamiltonian.
In fact, \cite{amin2015searching} suggests that the distribution of the samples
would correspond more closely to an instantaneous Hamiltonian at an intermediate
point in time, called the \emph{freeze-out point}. Therefore, our goal is to
investigate the applicability of sampling from a TFIM with significant $\Gamma$
to free energy--based reinforcement learning.

\subsection{Simulated Quantum Annealing}

Simulated quantum annealing (SQA) methods are a class of quantum-inspired
algorithms that perform discrete optimization by classically simulating
quantum tunnelling phenomena (see \cite[p. 422]{naturalcomputingalgorithms} for
an introduction). The algorithm used in this paper is a single spin-flip
version of quantum Monte Carlo numerical simulation based on the
Suzuki--Trotter formula, and uses the Metropolis acceptance probabilities.
The SQA algorithm simulates the quantum annealing phenomena of an Ising spin
model with a transverse field, that is,
\begin{equation}
\mathcal H(t) = - \sum_{i, j} J_{ij} \sigma^z_i \sigma^z_j 
- \sum_i h_i \sigma^z_i - \Gamma(t) \sum_i \sigma^x_i\,,
\end{equation}
where $\sigma^z$ and $\sigma^x$ represent the Pauli $z$- and $x$-matrices,
respectively, the indices $i$ and $j$ range over the sites of the system, and
the time $t$ ranges from $0$ to $1$. In this quantum evolution, the strength of
the transverse field is slowly reduced to zero at finite temperature. In our
implementations, we have used a linear transverse field schedule for the SQA
algorithm as in \cite{1411.5693} and \cite{martovnak2002quantum}. Based on the
Suzuki--Trotter formula, the key idea of this algorithm is to approximate the
partition function of the Ising model with a transverse field as a partition
function of a classical Hamiltonian denoted by $\mathcal H^\eff$, corresponding
to a classical Ising model of one dimension higher. More precisely,
\begin{equation}
\mathcal H^\eff(\mathbf{\sigma})=
-\sum_{i,j}\sum_{k=1}^{r}\frac{J_{ij}}{r}\sigma_{ik}\sigma_{jk}-J^+\sum_{i}
\sum_{k=1}^{r}\sigma_{ik}\sigma_{i,k+1}
-\sum_{i}\sum_{k=1}^{r}\frac{h_i}{r}\sigma_{ik}\,,
\end{equation}
where $r$ is the number of replicas, $J^+ = \frac{1}{2\beta} \log \coth
\left(\frac{\Gamma\beta}{r}\right)$, and $\sigma_{ik}$ represent spins of the
classical system of one dimension higher.

In our experiments, the strength $\Gamma$ of the transverse field is scheduled
to linearly decrease from $20.00$ to one of $\Gamma_f = 0.01$ or $2.00$. The
inverse temperature $\beta$ is set to the constant $2.00$. The initial value,
$20.00$, of the transverse field is empirically chosen to be well above the
coupling strengths created during the training. Each spin is replicated 25 times
to represent the Trotter slices in the extra dimension. The simulation is set to
iterate over all replications of all spins one time per sweep, and the number of
sweeps is set to 300, which appears to be large enough for the sizes of
Ising models constructed during our experiments. For each instance of input, the
SQA algorithm is run 150 times. After termination, the configuration of each
replica, as well as the configuration of the entire classical Ising model of one
dimension higher, is returned.

Although the SQA algorithm does not follow the dynamics of a physical quantum
annealer explicitly, it is used to simulate this process, as it captures major
quantum phenomena such as tunnelling and entanglement \cite{1510.08057}. In
~\cite{1510.08057}, for example, it is shown that quantum Monte Carlo
simulations can be used to understand the tunnelling behaviour in quantum
annealers. As mentioned previously, it readily follows from the results of
\cite{morita2006convergence} that the limiting distribution of SQA is the
Boltzmann distribution of $\mathcal H^{\eff}$. This makes SQA a candidate
classical algorithm for sampling from Boltzmann distributions of classical and
quantum Hamiltonians. The former is achieved by setting $\Gamma_f \simeq 0$, and
the latter by constructing an effective Hamiltonian of the system of one
dimension higher, representing the quantum Hamiltonian with non-negligible
$\Gamma_f$. Alternatively, a classical Monte Carlo simulation used to sample
from the Boltzmann distribution of the classical Ising Hamiltonian is the SA
algorithm, based on thermal fluctuations of classical spin systems.

\subsection{Markov Decision Process}\label{mdp}
The stochastic control problem of interest to us is a Markov decision process
(MDP), defined as having:
\begin{enumerate}[label=(\roman*), leftmargin=2pc, itemsep=0pt]
  \item finite sets of states $S$ and actions $A$;\footnote{\, When both $S$
and $A$ are finite, the MDP is said to be finite.}
  \item a controlled Markov chain \cite{Yuksel}, defined by a transition kernel
$\mathbb P( s' \in S | s \in S, a \in A)$;\footnote{\, The transition kernel
does not need to be time-homogeneous; however, this definition suffices for the
purposes of this work.}
  \item a real-valued function $r: S \times A \to \mathbb R$, known as the
\emph{immediate reward structure}; and
  \item a constant $\gamma \in [0, 1)$, known as the \emph{discount factor}.
\end{enumerate}

A function $\pi: S \to A$ is called a \emph{stationary policy}; that is, it is a
choice of action $\pi(s)$ for every state $s$ independent of the point in time
that the controlled process reaches $s$. The application of a stationary policy
$\pi$ reduces the MDP into a time-homogeneous Markov chain $\Pi$, with a
transition probability $\mathbb P (s' | s , \pi(s))$.\footnote{\, For 
more-general statements, see \cite{Yuksel}.} The random process $\Pi$ with
initial condition $\Pi_0 = s$ we denote by $\Pi^s$.

Our \emph{Markov decision problem} is to find
\begin{align}\label{optimization}
\pi^* (s) = \argmax_{\pi} V (\pi, s),
\end{align}
where
\begin{align}\label{value}
    V (\pi, s) = \mathbb{E} \left[ \sum\limits_{i=0}^{\infty} \gamma^{i}\,
    r(\Pi^s_i,\pi(\Pi^s_i)) \right].
\end{align}

\subsubsection{Maze Traversal as a Markov Decision Process}\label{maze-mdp}

Maze traversal is a problem typically used to develop and benchmark
reinforcement learning algorithms \cite{sutton1990integrated}. A maze is
structured as a two-dimensional grid of $r$ rows and $c$ columns in which a
decision-making agent is free to move up, down, left, or right, or to stand
still. During the maze traversal, the agent encounters obstacles (e.g., walls),
rewards (e.g., goals), and penalties (negative rewards, e.g., a pit). Each cell
of the maze can contain either a deterministic or stochastic reward, a wall, a
pit, or a neutral value. Fig.~\ref{fig:grid-det} and Fig. \ref{fig:grid-stoc}
show examples of two mazes. Fig.~\ref{fig:grid-sol} shows the corresponding
solutions to the maze in Fig.~\ref{fig:grid-det}.

The goal of the reinforcement learning algorithm in the maze traversal problem
is for the agent to \emph{learn} the optimal action to take in each cell of the
maze by maximizing the total reward, that is, finding a route across the maze
that avoids walls and pits while favouring rewards. This problem can be
modelled as an MDP determined by the following components:
\begin{itemize}[leftmargin=1pc, itemsep=-0pt]
  \item The state of the system is the agent's position within the maze. The
position state $s$ takes values in the set of states
  $$S = \{1,...,r\} \times \{1,...,c\}.$$
  \item In any state, the agent can decide to take one of the five actions
  $$a \in \{\uparrow, \downarrow, \leftarrow, \rightarrow,
\circlearrowleft\}.$$
   These actions will guide the agent through the maze. An action that would
lead the agent into a wall ($W$) or outside of the maze boundary is treated as
an inadmissible action.
   Each action can be viewed as an endomorphism on the set of states
   $$a: S \to S.$$
  If $a=$ $\circlearrowleft$, then $a(s) = s$; otherwise, $a (s)$ is the state
adjacent to $S$ in the direction shown by $a$. We do not consider training
samples where $a$ is inadmissible.
   \item \label{kernel} The transition kernel determines the probability of the
agent moving from one state to another given a particular choice of action. In
the simplest case, the probability of transition from $s$ to $a(s)$ is one:
\begin{align*}
\mathbb{P}(a(s)|s,a)=1.
\end{align*}
We call the maze \emph{clear} if the associated transition kernel is as above,
as opposed to the \emph{windy} maze, in which there is a nonzero probability
that if the action $a$ is taken at state $s$, the next state will differ from
$a(s)$.
  \item The immediate reward $r(s, a)$ that the agent gains from taking an
action $a$ in state $s$ is the value contained in the destination state. Moving
into a cell containing a reward returns the favourable value $R$, moving into a
cell containing a penalty returns the unfavourable value $P$, and moving into a
cell with no reward returns a neutral value in the interval $(P, R)$.
  \item A discount factor for future rewards is a non-negative constant $\gamma
<1$. In our experiments, this discount factor is set to \mbox{$\gamma = 0.8$}.
The discount factor is a feature of the problem rather than a free parameter of
an implementation. For example, in a financial application scenario, the
discount factor might be a function of the risk-free interest rate.
\end{itemize}

The immediate reward for moving into a cell with a stochastic reward is given by
a random variable $\mathcal R$. If an agent has prior knowledge of this
distribution, then it should be able to treat the cell as one with a
deterministic reward value of $\mathbb E[\mathcal R]$. This allows us to find
the set of all optimal policies in each maze instance. This policy information
is denoted by \mbox{$\alpha^\ast : S \to 2^A$}, associating with each state $s
\in S$ a set of optimal actions $\alpha^\ast (s) \subseteq A$.

In our maze model, the neutral value is set to 100, the reward $R = 200$, and
the penalty $P= 0$. In our experiments, the stochastic reward $\mathcal R$ is
simulated by drawing a sample from the Bernoulli distribution $200\,
\mathrm{Ber} (0.5)$; hence, it has the expected value $\mathbb E[\mathcal R]=
100$, which is identical to the neutral value. Therefore, the solutions depicted
in Fig.~\ref{fig:grid-sol} are solutions to the maze of Fig.~\ref{fig:grid-stoc}
as well.

\subsection{Value Iteration} Bellman \cite{bellman1956dynamic} writes $V(\pi,
s)$ recursively in the following manner using the monotone convergence theorem:
\begin{align*}
V(\pi, s)
&= \mathbb{E} \left[ \sum\limits_{i=0}^{\infty}
\gamma^{i}\,  r(\Pi^s_i,\pi(\Pi^s_i)) \right] \\
&= \mathbb E[ r(\Pi^s_0,\pi(\Pi^s_0))] + \gamma\,
\mathbb{E} \left[
\sum\limits_{i=0}^{\infty} \gamma^{i}\, r(\Pi^s_{i+1},\pi(\Pi^s_{i+1}))
\right] \\
&= \mathbb E[r(s,\pi(s)) ]+
\gamma \sum_{s' \in S} \mathbb P(s'|s, \pi(s))\, V(\pi, s')\,.
\end{align*}
In particular, it leads to the Bellman optimality equation:
\begin{equation}\label{value-iteration-on-V}
V^\ast(s) =
V(\pi^\ast, s)\\ = \max_a \left( \mathbb E[r(s,a)]
+ \gamma  \sum_{s' \in S} \mathbb P(s' |s, a)\, V^\ast(s')\right).
\end{equation}
Hence, $V^\ast$ is a fixed point for the operator $$T_V (f): s \mapsto \max_a
\left(\mathbb E[r (s,a)] + \gamma \int f\right)$$ on the space $L_\infty (S)$ of
bounded functions $S \to \mathbb R$ endowed with the max norm. Here, the
integral is taken with respect to the probability measure on $S$, induced by the
conditional probability distribution $\mathbb P(s'|s, a)$. It is easy to check
that $T_V$ is a contraction mapping, and thus $V^\ast$ is the unique fixed point
of $T_V$ and the uniform limit of any sequence of functions $\{ T^n_V f\}_{n}$.
Numerical computation of this limit using \eqref{value-iteration-on-V}, called
\emph{value iteration}, is a common method of solving the Markov decision
problem~(\ref{optimization}). However, even the \mbox{${\varepsilon}$-optimal}
algorithms for this approach depend heavily on the cardinality of both ${S}$ and
${A}$, and suffer from the \emph{curse of dimensionality}
\cite{bellman1956dynamic, putermanMDP}. Moreover, the value iteration method
requires having full knowledge of the transition probabilities, as well as the
distribution of the immediate rewards.

\subsection{Q-functions} \label{q-func}
For a stationary policy $\pi$, the \mbox{\emph{Q-function}} (also known as the
\emph{action--value function}) is defined as a mapping of a pair $(s, a)$ to the
expected value of the reward of the Markov chain that begins with taking action
$a$ at initial state $s$ and continuing according to $\pi$~\cite{sutton-book}:
\begin{align*}
Q (\pi, s, a) &= \mathbb E[r(s,a)] + \mathbb{E} \left[
\sum\limits_{i=1}^{\infty} \gamma^{i}\,
    r(\Pi^s_{i},\pi(\Pi^s_{i})) \right].
\end{align*}
It is straightforward to check that
\begin{align*}
V(\pi^\ast, s)= \max_a Q(\pi^\ast, s, a),
\end{align*}
and for $Q^\ast(s, a)= \max_\pi Q(\pi, s, a)=Q(\pi^\ast, s, a)$, the optimal
policy for the MDP can be retrieved via the following:
\begin{align}\label{eq:policy-from-Q}
\pi^\ast (s) = \argmax_a Q^\ast (s, a).
\end{align}
This reduces the Markov decision problem to computing $Q^\ast (s, a)$. The
Bellman optimality equation for $Q^\ast( s, a)$ is
\begin{equation*}
\!\!\!Q^\ast( s, a) = \mathbb E[r(s, a) ] + \gamma \sum_{s'} \mathbb P(s' |
s, a) \max_{a'} Q^\ast ( s', a'),
\end{equation*}
which makes $Q^\ast$ the fixed point of a different operator
$$T_Q(f): (s, a) \mapsto \mathbb E[r(s, a)] + \gamma \int \max_{a'} f$$
defined on $L_\infty (S \times A)$.

\subsection{Temporal-Difference Gradient Descent}\label{sec:td0}

In this section, we derive the \mbox{Q-learning} method for MDPs. From the
previous section, we know that starting from an initial $Q_0: S \times A \to
\mathbb R$, the sequence $\{ Q_n = T^n_Q Q_0\}$ converges to $Q^\ast$. The
difference
\begin{equation}\label{temp-diff}
	Q_{n+1}(s, a) -Q_n(s,a) = \mathbb E[r(s, a)] \\+ \gamma \sum_{s'}
\mathbb P(s' | s, a)\max\limits_a \underbrace{Q_n(s', a)}_{(\ast)} - Q_n (s, a)
\end{equation}
is called the \emph{temporal difference} of \mbox{Q-functions}, and is denoted
by $E_\mathrm{TD}$.

Employing a gradient approach to find the fixed point of $T$ on $L_\infty (S
\times A)$ involves locally parametrizing the functions in this space by a
vector of parameters $\boldsymbol{\theta}$, that~is,
$$Q(s, a) = Q(s, a; \boldsymbol{\theta}),$$
and travelling in the direction that minimizes~$\|E_\mathrm{TD}\|^2$:
\begin{align}\label{descent-direction}
\Delta \boldsymbol{\theta} \propto - E_\mathrm{TD} \nabla_{\boldsymbol{\theta}}
E_\mathrm{TD}\,.
\end{align}

\def\app#1#2{%
  \mathrel{%
    \setbox0=\hbox{$#1\sim$}%
    \setbox2=\hbox{%
      \rlap{\hbox{$#1\propto$}}%
      \lower1.2\ht0\box0%
    }%
    \raise0.27\ht2\box2%
  }%
}
\def\approxprop{\mathpalette\app\relax}

The method TD(0) consists of treating the $(\ast)$ in \eqref{temp-diff} as
constant with respect to the parameterization $\boldsymbol{\theta}$, in which
case we may write
$$\Delta \boldsymbol{\theta} \approxprop
E_\mathrm{TD}(s,a)\nabla_{\boldsymbol{\theta}} Q(s,a;\boldsymbol{\theta}).$$

For an agent agnostic with respect to the transition kernel or
the distribution of the reward $r(s, a)$, or both, this update rule for
$\boldsymbol{\theta}$ is not possible. The alternative is to substitute, at each
iteration, the expected value
$$\sum_{s'} \mathbb P(s' | s, a)\max\limits_a Q_n(s_{n+1}, a)$$
by $\max_a Q_n (s_{n+1}, a)$, where $s_{n+1}$ is drawn from the probability
distribution $\mathbb P(s'|s, a)$, and substitute $\mathbb E[r(s_n, a_n)]$ by a
sample of $r(s_n, a_n)$. This leads to a successful Monte Carlo training method
called \mbox{\emph{Q-learning}}.

In what follows, we explain the case in which $\boldsymbol{\theta}$ comprises
the weights of a Boltzmann machine. Let us begin by introducing clamped
Boltzmann machines, which are of particular importance in the case of
reinforcement learning.

\subsection{Clamped Boltzmann Machines}\label{sec:boltzmann}

A classical Boltzmann machine is a type of stochastic neural network with two
sets $V$ and $H$ of visible and hidden nodes, respectively. Both visible and
hidden nodes represent binary random variables. We use the same notation for a
node and the binary random variable it represents. The interactions between the
variables represented by their respective nodes are specified by real-valued
weighted edges of the underlying undirected graph. A GBM, as opposed to models
such as RBMs and DBMs, allows weights between any two nodes.

The \emph{energy} of the classical Boltzmann machine is
\begin{equation}
\mathscr{E}(\mathbf v, \mathbf h)= -\sum_{v \in V,\, h \in H} w^{vh}v h
-\sum_{\{v, v'\} \subseteq V } w^{vv'} vv' -\sum_{\{h, h'\} \subseteq H}
w^{hh'} hh',
\end{equation}
with $w^{vh}$, $w^{vv'}$, and $w^{hh'}$ denoting the weights between
visible and hidden, visible and visible, and hidden and hidden
nodes of the Boltzmann machine, respectively, defined
as a function of binary vectors $\bf v$ and $\bf h$ corresponding to the
visible and hidden variables, respectively.

A clamped GBM is a neural network whose underlying graph is the subgraph
obtained by removing the visible nodes for which the effect of a fixed
assignment $\mathbf v$ of the visible binary variables contributes as constant
coefficients to the associated energy
\begin{equation}
\mathscr{E}_{\mathbf v}(\mathbf h)= -\sum_{v \in V,\, h \in H} w^{vh}v h
-\sum_{\{v, v'\} \subseteq V } w^{vv'} vv' -\sum_{\{h, h'\} \subseteq H}
w^{hh'} hh'\,.   \label{GBMh}
\end{equation}

A clamped quantum Boltzmann machine (QBM) has the same underlying graph as a
clamped GBM, but instead of a binary random variable, a qubit is associated to
each node of the network. The energy function is substituted by the quantum
Hamiltonian
\begin{equation}\label{qbm-hamiltonian}
\mathcal{H}_{\mathbf v}= -\sum_{v \in V,\, h \in H} w^{vh}v \sigma^z_h
-\sum_{\{v, v'\} \subseteq V } w^{vv'} vv' -\sum_{\{h, h'\} \subseteq H}
w^{hh'} \sigma^z_h\sigma^z_{h'} - \Gamma \sum_{h \in H} \sigma_h^x\,,
\end{equation}
where $\sigma_h^z$ represent the Pauli $z$-matrices and $\sigma^x_h$ represent
the Pauli $x$-matrices.
Thus, a clamped QBM with $\Gamma = 0$ is equivalent to a clamped classical
Boltzmann machine. This is because $\mathcal H_{\mathbf v}$ is a diagonal matrix
in the $\sigma^z$-basis, the spectrum of which is identical to the range of
$\mathscr E_{\mathbf v}$. The remainder of this section is formulated for the
clamped QBMs, acknowledging that it can easily be specialized to clamped
classical Boltzmann machines.

Let $\beta= \frac{1}{k_B T}$ be a fixed thermodynamic beta. For an assignment
of visible variables $\mathbf v$, $F(\mathbf{v})$ denotes the \emph{equilibrium
free energy}, and is defined as
\begin{align}\label{eq:free-def}
F(\mathbf{v}) := - \frac{1}\beta \ln Z_{\mathbf v} = \langle \mathcal H_{\mathbf v}
\rangle + \frac{1}\beta \tr (\rho_{\mathbf v} \ln \rho_{\mathbf v})\,.
\end{align}
Here, $Z_{\mathbf v} = \tr (e^{-\beta \mathcal H_{\mathbf v}})$ is the partition
function of the clamped QBM and $\rho_{\mathbf v}$ is the density matrix
$\rho_{\mathbf v} = \frac{1}{Z_{\mathbf v}}e^{-\beta \mathcal H_{\mathbf v}}$. The
term $- \tr (\rho_{\mathbf v} \ln \rho_{\mathbf v})$ is the entropy of the system.
The notation $\langle \cdots \rangle$ is used for the expected value of any
observable with respect to the Gibbs measure, in particular,
$$\langle \mathcal H_{\mathbf v} \rangle = \frac{1}{Z_{\mathbf v}} \tr( \mathcal
H_{\mathbf v} e^{-\beta \mathcal H_{\mathbf v}}).$$

\subsection{Reinforcement Learning Using Clamped Boltzmann Machines} \label{RL}

In this section, we explain how a general Boltzmann machine (GBM) can be used
to provide a Q-function approximator in a Q-learning method. To the best of our
knowledge, this derivation has not been previously given, although it can be
readily derived from the ideas presented in \cite{HintonRBM} and
\cite{1601.02036}.
Following \cite{HintonRBM}, the goal is to use the negative free energy of a
Boltzmann machine to approximate the \mbox{Q-function} through the relationship
$$Q (s, a) \approx -F (\mathbf s, \mathbf a) = - F (\mathbf s, \mathbf a;
\boldsymbol \theta)$$
for each admissible state--action pair \mbox{$(s, a) \in S\times A$}. Here,
$\mathbf s$ and $\mathbf a$ are binary vectors encoding the state $s$ and
action $a$ on the state nodes and action nodes, respectively, of the Boltzmann
machine. In reinforcement learning, the visible nodes of the GBM are
partitioned into two subsets of state nodes $S$ and action nodes $A$.

The parameters $\boldsymbol \theta$, to be trained according to a TD(0) update
rule (see Sec. \ref{sec:td0}), are the weights in a Boltzmann machine. For
every weight $w$, the update rule is
$$\Delta w =
-\varepsilon (r_n({s}_n, {a}_n)
+\gamma \max_a Q ({s}_{n+1}, a) - Q({s}_n,{a}_n) ) \frac{\partial F}{\partial
w}\,.$$

From \eqref{eq:free-def}, we obtain
\begin{align*}
\frac{\partial F(\mathbf s, \mathbf a)}{\partial w} &=  - \frac{1}{\beta Z_
{\mathbf s,
\mathbf a}} \frac{\partial}{\partial w} \tr \left(e^{-\beta \mathcal H_{\mathbf
s,\mathbf a}}\right)\\
&=  \frac{1}{\beta Z_{\mathbf s, \mathbf a}} \tr \left(\beta e^{-\beta \mathcal
H_{\mathbf s, \mathbf a}} \frac{\partial}{\partial w} \mathcal H_{\mathbf s,
\mathbf a} \right) \\
&=  \left\langle \frac{\partial}{\partial w} \mathcal H_{\mathbf s,
\mathbf a} \right\rangle.
\end{align*}
Therefore, the update rule for TD(0) for the clamped QBM can be rewritten as
\begin{equation}\label{q-vh-update}
\Delta w^{vh} =\,
\varepsilon  (r_n({s}_n, {a}_n)
+\gamma Q ({s}_{n+1}, {a}_{n+1}) - Q({s}_n,{a}_n) ) v \langle \sigma_h^z
\rangle
\end{equation}
and
\begin{equation}\label{q-hh-update}
\Delta w^{hh'} =\,
\varepsilon (r_n({s}_n, {a}_n)
+\gamma Q ({s}_{n+1}, {a}_{n+1}) - Q({s}_n,{a}_n) ) \langle
\sigma_h^z\sigma_{h'}^z \rangle,
\end{equation}
where the thermodynamic beta is absorbed into the learning rate $\varepsilon$,
and
$$a_{n+1} = \argmax_a Q(s_{n+1}, a).$$
Here, $h$ and $h'$ denote two distinct hidden nodes and (by a slight abuse of
notation) the letter $v$ stands for a visible (state or action) node, and also
the value of the variable associated to that node.

To approximate the right-hand side of each of \eqref{q-vh-update} and
\eqref{q-hh-update}, we use SQA experiments. By \cite[Theorem
6]{suzuki1976relationship}, we may find the expected values of the observables
$\langle\sigma_h^z\rangle$ and $\langle \sigma_h^z \sigma_{h'}^z \rangle$ by
averaging the corresponding spins in the classical Ising model of one dimension
higher used in SQA. To approximate the Q-function, we take advantage of
\cite[Theorem 4]{suzuki1976relationship} and use \eqref{eq:free-def} applied to
this classical Ising model. More precisely, let $\mathcal H^\eff_{\mathbf v}$
represent the Hamiltonian of the classical Ising model of one dimension higher
and the associated energy function $\mathscr E^\eff_{\mathbf v}$. The free energy
of this model can be written
\begin{align}
\label{freeen-eff}
F (\mathbf{v}) = \langle \mathcal H^\eff_{\mathbf v} \rangle +
\frac{1}\beta
\sum_{c}\mathbb{P}(c|\mathbf{v}) \log \mathbb{P}(c|\mathbf{v})\,,
\end{align}
where $c$ ranges over all spin configurations of the classical Ising model of
one dimension higher.

The above argument holds in the absence of the transverse field, that is, for
the classical Boltzmann machine. In this case, the TD(0) update rule is given by
\begin{equation}\label{vh-update}
\Delta w^{vh} =\,
\varepsilon  (r_n({s}_n, {a}_n) 
+\gamma Q ({s}_{n+1}, {a}_{n+1}) - Q({s}_n,{a}_n) ) v \langle h \rangle
\end{equation}
and
\begin{equation}\label{hh-update}
\Delta w^{hh'} =\,
\varepsilon (r_n({s}_n, {a}_n)
+\gamma Q ({s}_{n+1}, {a}_{n+1}) - Q({s}_n,{a}_n) ) \langle hh' \rangle\,,
\end{equation}
where $\langle h \rangle$ (referred to as \emph{activations} of the hidden
nodes in machine learning terminology) and $\langle h h' \rangle$ are the
expected values of the variables and the product of variables, respectively, in
the binary encoding of the hidden nodes with respect to the Boltzmann
distribution given by
$\mathbb{P}(\mathbf{h}|\mathbf{v}) = {\exp(-\beta\mathscr{E}_{\mathbf{v}}
(\mathbf{h}))}/{\sum_{{\mathbf{h}}'}\exp(-\beta\mathscr{E}_{\mathbf{v}}
({\mathbf{h}}'))}
$. Therefore, they may be approximated using SA or SQA when $\Gamma \to 0$.

The values of the Q-functions in \eqref{vh-update} and \eqref{hh-update} can
also be approximated empirically, since, in a classical Boltzmann machine,
\begin{align}
\label{freeen}
&F (\mathbf{v}) = \sum_{\mathbf{h}}\mathbb{P}(\mathbf{h}|\mathbf{v}) \mathscr{E}_{\mathbf
v} (\mathbf{h}) + \frac{1}\beta
\sum_{\mathbf{h}}\mathbb{P}(\mathbf{h}|\mathbf{v}) \log
\mathbb{P}(\mathbf{h}|\mathbf{v})\\
&\quad= - \sum_{\substack{s \in S\\h \in H}}w^{sh}s \langle h\rangle -
\sum_{\substack{a \in A\\h \in H}} w^{ah}a \langle h\rangle
- \sum_{\{h, h'\} \subseteq H} u^{hh'}\langle hh'\rangle + \frac{1}\beta
\sum_{\mathbf{h}}\mathbb{P}(\mathbf{h}|\mathbf{s}, \mathbf{a}) \log
\mathbb{P}(\mathbf{h}|\mathbf{s}, \mathbf{a}). \nonumber
\end{align}

\begin{rmk}\label{rmk:rbm}
In the case of an RBM, Sallans and Hinton \cite{HintonRBM} show that the free
energy is given by
\begin{equation}\label{RFREE}
- F(\mathbf{s},\mathbf{a}) = \sum_{\substack{s \in S\\h \in H}}w^{sh}s \langle
h\rangle+ \sum_{\substack{a \in A\\h \in H}} w^{ah}a \langle h\rangle
		 - \frac{1}{\beta}\sum_{h \in H} \left[ \langle h\rangle
\log\langle h\rangle + (1-\langle h\rangle)\log(1-\langle h\rangle)\right].
\end{equation}
The update rule for the weights of the RBM is \eqref{vh-update} alone.
Moreover, in the case of RBMs, the equilibrium free energy $F(\mathbf{s,a})$ and
its derivatives with respect to the weights can be calculated without the need
for Boltzmann distribution sampling, according to the closed formula
\begin{align}\label{eq:rbm-closed-expected-value}
\langle h\rangle &= \mathbb{P}(\sigma_h = 1 | \mathbf{s,a}) = \sigma
\left(\sum_{s \in S} w^{sh}s+ \sum_{a \in A} w^{ah}a\right)  \\
&= \left\{1 + \text{exp}\left(- \sum_{s \in S} w^{sh}s - \sum_{a \in A}
w^{ah}a\right)\right\}^{-1}\,.\nonumber
\end{align}
Here, $\sigma$ denotes the sigmoid function. Note that, in the general case,
since the hidden nodes of a clamped Boltzmann machine are not independent, the
calculation of the free energy is intractable.
\end{rmk}

\begin{algorithm}[t]
    \caption{\, RBM-RL}
    \label{alg:RL-RBM}
    \begin{algorithmic}[1]
        \State initialize weights of RBM
    \ForAll {training samples $(s_1, a_1)$}
                 \State $s_2 \gets a_1 (s_1)$, $a_2 \gets \argmax_a Q(s_2, a)$
                 \State calculate $\langle {h}_i \rangle \text{ for }(i=1, 2)$
using \eqref{eq:rbm-closed-expected-value}
                 \State calculate $F(\mathbf s_i, \mathbf a_i) \text{ for
}(i=1, 2)$ using \eqref{RFREE}
        \State $Q(s_i, a_i) \gets -F (\mathbf s_i, \mathbf a_i)$ for $(i=1, 2)$
        \State update RBM weights using (\ref{ws}) and (\ref{wa})
        \State $\pi(s_1) \gets \argmax_a Q (s_1, a)$
    \EndFor
        \State \textbf{return} $\pi$
    \end{algorithmic}
\end{algorithm}

\section{Algorithms} \label{sec:algorithms}

In this section, we present the details of classical reinforcement learning
using RBM, a semi-classical approach based on a DBM (using SA and SQA), and a
quantum reinforcement learning approach (using SQA or quantum annealing). All 
of the algorithms are based on the Q-learning TD(0) method presented in the
previous section. Pseudo-code for these methods is provided in Algorithms
\ref{alg:RL-RBM}, \ref{alg:CRL-DBM}, and \ref{alg:QBM-RL} below.

\subsection{Reinforcement Learning Using RBMs}
The RBM reinforcement learning algorithm is due to Sallans and Hinton
\cite{HintonRBM}. This algorithm uses the update rule (\ref{vh-update}), with
$v$ representing state or action encoding, to update the weights of an RBM, and
(\ref{eq:rbm-closed-expected-value}) to calculate the expected values of random
variables associated with the hidden nodes $\langle h \rangle$. As explained in
Sec. \ref{RL}, the main advantage of RBM is that it has explicit formulas for
the hidden-node activations, given the values of the visible nodes. Moreover,
only for RBMs can the entropy portion of the free energy \eqref{freeen} be
written in terms of the activations of the hidden nodes. More-complicated
network architectures do not possess this property, so there is a need for a
Boltzmann distribution sampler.

In Algorithm \ref{alg:RL-RBM}, we recall the steps of the classical
reinforcement learning algorithm using an RBM with a graphical model similar to
that shown in Fig.~\ref{rbm-graph}. We set the initial Boltzmann machine
weights using Gaussian zero-mean values with a standard deviation of $1.00$, as
is common practice for implementing Boltzmann machines~
\cite{hinton2010practical}.
Consequently, this initializes an approximation of a Q-function and a policy
$\pi$ given by
$$\pi (s) = \argmax_a Q(s, a)\,.$$
In each training iteration, we select a state--action pair $(s_1, a_1) \in S
\times A$. We associate a classical spin variable $\sigma_h$ to each hidden
node $h$. Then, the activations of the hidden nodes are calculated via
\eqref{eq:rbm-closed-expected-value}.
In our experiments, all Boltzmann machines have as many state nodes as $|S|$
and as many action nodes as $|A|$. We associate one node for every state $s \in
S$, and the corresponding binary encoding is $\mathbf s = (0, 0, \ldots, 1,
\ldots, 0)$, with zeroes everywhere except at the index of the node
corresponding to $s$. We use similar encoding for the actions, using the
action nodes. A subsequent state $s_2$ is obtained from the state--action pair
$(s_1, a_1)$ using the transition kernel outlined in Sec.
\ref{sec:preliminaries}, and a corresponding action $a_2$ is chosen via policy
$\pi$. The free energy of the RBM is calculated using \eqref{RFREE} for both
$(s_1, a_1)$ and $(s_2, a_2)$.

This results in an approximation of the Q-function (see Sec. \ref{q-func})
defined on the state--action space $S \times A$ \cite{HintonRBM},
$$Q ({s}, {a}) \approx -F (\mathbf s, \mathbf a)\,,$$
for both state--action pairs.
We then use the update rule \eqref{vh-update}, or, more precisely,
\begin{equation}\label{ws}
\Delta w^{sh} =
\varepsilon  (r({s}_1, {a}_1)
+\gamma Q ({s}_{2}, {a}_{2}) - Q({s}_1,{a}_1) ) s_1 \langle h \rangle
\end{equation}
and
\begin{equation}\label{wa}
\Delta w^{ah} =
\varepsilon ( r({s}_1, {a}_1)
+\gamma Q ({s}_{2}, {a}_{2}) - Q({s}_1,{a}_1)) a_1 \langle h \rangle \, ,
\end{equation}
with a learning rate $\varepsilon$ to update the weights of the RBM. In view of
\eqref{eq:policy-from-Q}, the best known policy can be acquired via \mbox{$\pi
(s) = \argmax_a Q(s, a)$} for any state $s$.

\begin{algorithm}[t]
    \caption{\, DBM-RL}
    \label{alg:CRL-DBM}
    \begin{algorithmic}[1]
        \State initialize weights of DBM
    \ForAll {training samples $(s_1, a_1)$}
                 \State $s_2 \gets a_1 (s_1)$, $a_2 \gets \argmax_a Q(s_2, a)$
                \State approximate $\langle {h}_i \rangle, \langle {h}_i {h'}_i
\rangle, \mathbb P(\mathbf h|\mathbf s_i, \mathbf a_i)$
\label{clr-dbm-approximate-sqa}
        \Statex \quad \, using SA or SQA for $(i=1, 2)$
        \State calculate $F(\mathbf s_i, \mathbf a_i)$ using (\ref{freeen}) for
($i = 1,2$)\label{clr-dbm-calcF-sqa}
        \State $Q(s_i, a_i) \gets -F (\mathbf s_i, \mathbf a_i)$ for $(i=1, 2)$
        \State update DBM weights using (\ref{hh-update}), (\ref{ws}), and 
        (\ref{wa}) \label{clr-dbm-update}
        \State $\pi(s_1) \gets \argmax_a Q (s_1, a)$
    \EndFor
        \State \textbf{return} $\pi$
    \end{algorithmic}
\end{algorithm}

\subsection{Reinforcement Learning Using DBMs}

Since we are interested in the dependencies between states and actions, we
consider a DBM architecture that has a layer of states connected to the first
layer of hidden nodes, followed by multiple hidden layers, and a layer of
actions connected to the final layer of hidden nodes (see Fig.~\ref{dbmgraph}).
We demonstrate the advantages of this deep architecture trained using SQA and
the derivation in Sec. \ref{RL} of the temporal-difference gradient method for
reinforcement learning using general Boltzmann machines (GBM).

In Algorithm \ref{alg:CRL-DBM}, we summarize the DBM-RL method. Here, the
graphical model of the Boltzmann machine is similar to that shown in Fig.~\ref
{dbm-graph}. The initialization of the weights of the DBM is performed in a
similar fashion to the previous algorithm.

In each training iteration, we select a state--action pair $(s_1, a_1) \in S
\times A$. Every node corresponding to a state or an action is removed from the
graph and the configurations of the spins corresponding to the hidden nodes are
sampled using SA or SQA on an Ising spin model constructed as follows: the
state $s_1$ contributes to a bias of $w^{s_1h}$ to $\sigma_h$ if $h$ is
adjacent to $s_1$; and the action $a_1$ contributes to a bias of $w^{a_1h}$ to
$\sigma_h$ if $h$ is adjacent to $a_1$. The bias on any spin $\sigma_h$ for
which $h$ is a hidden node not adjacent to state $s_1$ or action $a_1$ is zero.

A subsequent state $s_2$ is obtained from the state--action pair $(s_1, a_1)$
using the transition kernel outlined in Sec.~\ref{sec:preliminaries}, and a
corresponding action $a_2$ is chosen via policy $\pi$. Another SQA sampling is
performed in a similar fashion to the above for this pair.

According to lines \ref{clr-dbm-approximate-sqa} and \ref{clr-dbm-calcF-sqa} of
Algorithm \ref{alg:CRL-DBM}, the samples from the SA or SQA algorithm are used
to approximate the free energy of the classical DBM at points $(s_1, a_1)$ and
$(s_2, a_2)$ using \eqref{freeen}.

If SQA is used, averages are taken over each replica of each run; hence, there
are 3750 samples of configurations of the hidden nodes for each state--action
pair. The strength $\Gamma$ of the transverse field is scheduled to linearly
decrease from $20.00$ to $\Gamma_f = 0.01$.

The SA algorithm is used with a linear inverse temperature schedule that
increases from $0.01$ to $2.00$ in 50,000 sweeps, and is run 150 times. So, if
SA is used, there are only 150 sample points used in the above approximation.
The results of DBM-RL using SA or SQA have no significant differences.

The final difference between Algorithm~\ref{alg:RL-RBM} and Algorithm~
\ref{alg:CRL-DBM} is that the update rule now includes updates of weights
between two hidden nodes given by \eqref{hh-update},
\begin{equation}\label{wh}
\Delta w^{hh'} =
\varepsilon ( r({s}_1, {a}_1)
+\gamma Q ({s}_{2}, {a}_{2}) - Q({s}_1,{a}_1) ) \langle hh' \rangle\,, 
\end{equation}
in addition to the previous rules \eqref{ws} and \eqref{wa}.

\begin{algorithm}[b]
    \caption{\, QBM-RL}
    \label{alg:QBM-RL}
    \begin{algorithmic}[1]
        \State initialize weights of QBM
    \ForAll {training samples $(s_1, a_1)$}
                 \State $s_2 \gets a_1 (s_1)$, $a_2 \gets \argmax_a Q(s_2, a)$
        \State approximate $\langle {h}_i \rangle, \langle {h}_i {h'}_i
\rangle, \langle \mathcal H^\eff_{\mathbf{s_i},\mathbf{a_i}} \rangle,$
\label{q-approximate-sqaq}
        \Statex \quad \, and $\mathbb{P}(c|\mathbf{s_i},\mathbf{a_i})$ using SQA
for ($i = 1,2$) \nonumber
        \State calculate $F(\mathbf s_i, \mathbf a_i)$ using (\ref{freeen-eff})
for ($i = 1,2$)\label{q-calcF-sqaq}
        \State $Q(s_i, a_i) \gets -F (\mathbf s_i, \mathbf a_i)$ for $(i=1, 2)$
        \State update QBM weights using (\ref{hh-update}), (\ref{ws}), and
        (\ref{wa}) \label{q-update}
        \State $\pi(s_1) \gets \argmax_a Q (s_1, a)$
    \EndFor
        \State \textbf{return} $\pi$
    \end{algorithmic}
\end{algorithm}

\subsection{Reinforcement Learning Using QBMs}
The last algorithm is QBM-RL, presented in Algorithm \ref{alg:QBM-RL}. The
initialization is performed as in Algorithms \ref{alg:RL-RBM} and~
\ref{alg:CRL-DBM}. However, according to lines \ref{q-approximate-sqaq} and
\ref{q-calcF-sqaq}, the samples from the SQA algorithm are used to approximate
the free energy of a QBM at points $(s_1, a_1)$ and $(s_2, a_2)$ by computing
the free energy corresponding to an effective classical Ising spin model of one
dimension higher representing the quantum Ising spin model of the QBM, via
\eqref{freeen-eff}.

In this case, $\langle \mathcal H^\eff_{\mathbf{s},\mathbf{a}} \rangle$ from
\eqref{freeen-eff} is approximated by the average energy of the entire system
of one dimension higher and $\mathbb{P}(c|\mathbf{s},\mathbf{a})$ is approximated
by the normalized frequency of the configuration $c$ of the entire system of
one dimension higher (hence, there are only 150 sample points for each input
instance in this case). The strength $\Gamma$ of the transverse field in SQA is
scheduled to linearly decrease from $20.00$ to $\Gamma_f = 2.00$.
In this algorithm, the weights are updated as in Algorithm \ref{alg:CRL-DBM}.
However, $\langle h \rangle$ and $\langle h h' \rangle$ in this algorithm
represent expectations of measurements in the $z$-basis.

In each training iteration, we select a state--action pair $(s_1, a_1) \in S
\times A$. Every node corresponding to a state or an action is removed from
this graph and the configurations of the spins corresponding to the hidden
nodes are sampled using SQA on an Ising spin model constructed as follows: the
state $s_1$ contributes to a bias of $w^{s_1h}$ to $\sigma_h$ if $h$ is
adjacent to $s_1$; and the action $a_1$ contributes to a bias of $w^{a_1h}$ to
$\sigma_h$ if $h$ is adjacent to $a_1$. The bias on any spin $\sigma_h$ for
which $h$ is a hidden node not adjacent to state $s_1$ or action $a_1$ is zero.

A subsequent state $s_2$ is obtained from the state--action pair $(s_1, a_1)$
using the transition kernel outlined in Sec.~\ref{sec:preliminaries}, and a
corresponding action $a_2$ is chosen via policy $\pi$. Another SQA sampling is
performed in a similar fashion to the above for this pair.

In Fig.~\ref{1R1W1P} and Fig.~\ref{3R1W1P}, the selection of $(s_1, a_1)$ is
performed by sweeping across the set of state--action pairs. In
Fig.~\ref{fig:windb}, the selection of $(s_1, a_1)$ and $s_2$ is performed by
sweeping over $S \times A \times S$. In Fig.~\ref{smooth:1R1W1P}, the selection
of $s_1$, $a_1$, and $s_2$ are all performed uniformly randomly.

We experiment with a variety of learning-rate schedules, including exponential,
harmonic, and linear; however, we found that for the training of both RBMs and
DBMs, an adaptive learning-rate schedule performed best (for information on
adaptive subgradient methods, see \cite{duchi2011adaptive}). In our
experiments, the initial learning rate is set to 0.01.

In all of our studied algorithms, training terminates when a desired number of
training samples have been processed, after which the updated policy is
returned.

\section{Numerical Results} \label{results}

We study the performance of temporal-difference reinforcement learning
algorithms (explained in detail in Sec.~\ref{sec:algorithms}) using Boltzmann
machines. We generalize the method introduced in \cite{HintonRBM}, and compare
the policies obtained from these algorithms to the optimal policy using a
fidelity measure, which we define in \eqref{similaritymeasure}.

For $T_r$ independent trials of the same reinforcement learning
algorithm, $T_s$ training samples are used for reinforcement learning. The
fidelity measure at the $i$-th training sample is defined by
\begin{equation}
\text{fid}(i) =(T_r \times |S|)^{-1} \sum_{l = 1}^{T_r} \sum_{s\in S}
\mathbf{1}_{A(s, i, l) \in \alpha^*(s)},
\label{similaritymeasure}
\end{equation}
where $A(s, i, l)$ denotes the action assigned at the $l$-th run and $i$-th
training sample to the state $s$. In our experiments, each algorithm is run
1440 times, and for each run of an algorithm, $T_s= 500$ training samples are
generated.

\begin{figure*}[htbp]
\centering
\hspace{-.5cm}\subfloat[]{
\includegraphics[scale=0.65]{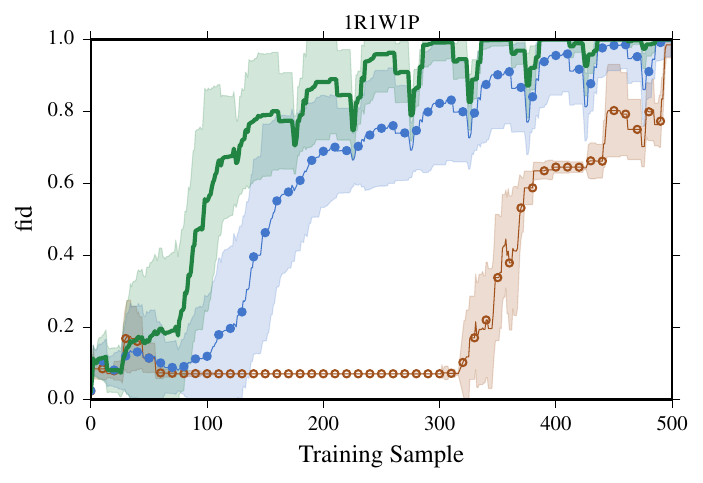}
\label{1R1W1P}}
\subfloat[]{
\includegraphics[scale=0.65]{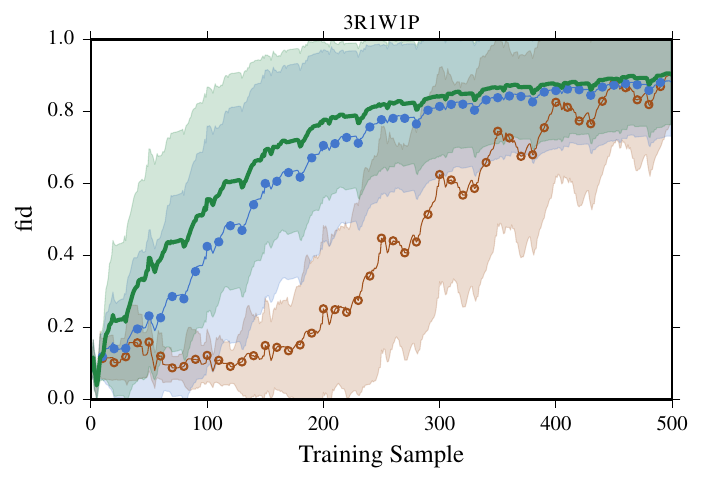}
\label{3R1W1P}}\\
\hspace{-.5cm}
\subfloat[]{
\includegraphics[scale=0.65]{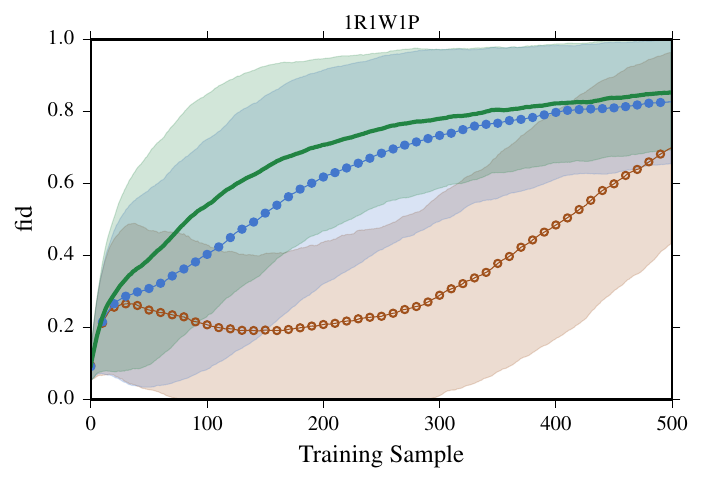}
\label{smooth:1R1W1P}}
\subfloat[]{
\includegraphics[scale=0.65]{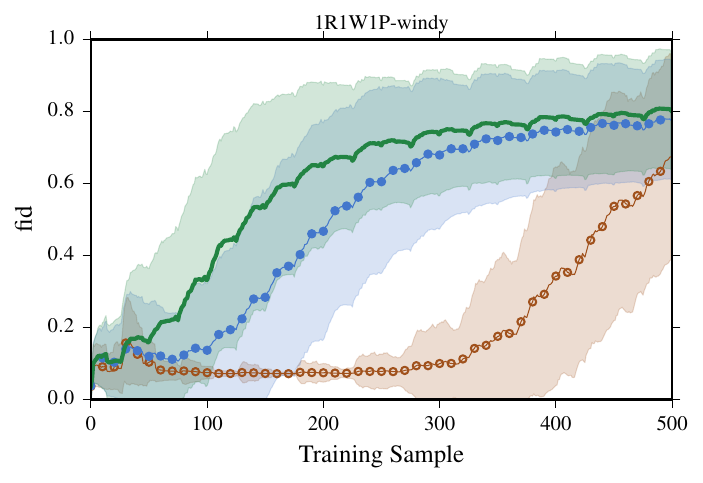}
\label{fig:windb}}\\
\subfloat{
    \includegraphics[scale=0.7]{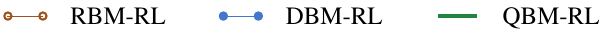}}
\vspace*{13pt}
\captionof{figure}{Comparison of RBM-RL, DBM-RL, and QBM-RL training results.
Every underlying RBM has 16 hidden nodes and every DBM has two layers of eight
hidden nodes. The shaded areas indicate the standard deviation of each training
algorithm.
 (a) The fidelity curves for the three algorithms run on the maze in Fig
\ref{fig:grid-det}. (b) The fidelity curves for the maze in Fig
\ref{fig:grid-stoc}. (c) The fidelity curves of the mentioned three algorithms
corresponding to the same experiment as that of (a), except that the training
is performed by uniformly generated training samples rather than sweeping across
the maze. (d) The fidelity curves corresponding to a windy maze similar to Fig
\ref{fig:grid-det}.
\label{fig:3by5}}
\end{figure*}

Fig.~\ref{1R1W1P} and Fig.~\ref{3R1W1P} show the fidelity of the generated
policies obtained from various reinforcement learning experiments on two clear
$3 \times 5$ mazes. In Fig.~\ref{1R1W1P}, the maze includes one reward, one
wall, and one pit, and in Fig.~\ref{3R1W1P}, the maze additionally includes two
stochastic rewards. In these experiments, the training samples are generated by
sweeping over the maze. Each sweep iterates over the maze elements in the same
order. This explains the periodic behaviour of the fidelity curves (cf.
Fig.~\ref{smooth:1R1W1P}).

The curves labelled `QBM-RL' represent the fidelity of reinforcement learning
using QBMs. Sampling from the QBM is performed using SQA. All other experiments
use classical Boltzmann machines as their graphical model. In the experiment
labelled `RBM-RL', the graphical model is an RBM, trained classically using
formula \eqref{eq:rbm-closed-expected-value}. The remaining curve is labelled
\mbox{`DBM-RL'} for classical reinforcement learning using a DBM. In these
experiments, sampling from configurations of the DBM is performed with SQA (with
$\Gamma_f = 0.01$). The fidelity results of DBM-RL coincide closely with those
of sampling configurations of the DBM using SA; therefore, we have not included
them. Fig.~\ref{smooth:1R1W1P} regenerates the results of Fig.~\ref{1R1W1P}
using uniform random sampling (i.e., without sweeping through the maze).

\begin{figure*}[htbp]
\centering
\subfloat[]{
\begin{tikzpicture}[scale=0.7]
\draw[ultra thick] (0,0) rectangle (5,-6);
\foreach \row in {0,...,5}{
  \foreach \column in {0,...,6} {
    \draw[dotted] (0,0) rectangle +(1*\row, -1*\column);}}
\draw[white, ultra thick] (0,-3) -- (0 ,-4);
\draw[white, ultra thick] (1,-3) -- (1 ,-4);
\draw[white, ultra thick] (2,-3) -- (2 ,-4);
\draw[white, ultra thick] (3,-3) -- (3 ,-4);
\draw[white, ultra thick] (4,-3) -- (4 ,-4);
\draw[white, ultra thick] (5,-3) -- (5 ,-4);
\node[align=left] at (0.5,-0.5) {$R$};
\node[align=left] at (4.5,-0.5) {$\mathcal{R}$};
\node[align=left] at (2.5,-1.5) {$W$};
\node[align=left] at (2.5,-2.5) {$W$};
\node[align=left] at (2.5,-4.5) {$W$};
\node[align=left] at (2.5,-3.3) {$\vdots$};
\node[align=left] at (0.5,-2.5) {$\mathcal{R}$};
\node[align=left] at (2.5, -5.5) {$P$};
\draw[ultra thick] (0,-7) rectangle (5,-12);
\foreach \row in {0,...,5}{
  \foreach \column in {0,...,5} {
    \draw[dotted] (0,-7) rectangle +(1*\row, -1*\column);}}
\draw[white, ultra thick] (0,-9) -- (0 ,-10);
\draw[white, ultra thick] (1,-9) -- (1 ,-10);
\draw[white, ultra thick] (2,-9) -- (2 ,-10);
\draw[white, ultra thick] (3,-9) -- (3 ,-10);
\draw[white, ultra thick] (4,-9) -- (4 ,-10);
\draw[white, ultra thick] (5,-9) -- (5 ,-10);
\node[align=left] at (0.5,-7.5) {$\circlearrowleft$};
\node[align=left] at (2.5,-8.5) {$W$};
\node[align=left] at (2.5,-10.5) {$W$};
\node[align=left] at (2.5,-9.3) {$\vdots$};
\node[align=left] at (0.5, -11.5) {$\uparrow$};
\node[align=left] at (4.5, -7.5) {$\leftarrow$};
\node[align=left] at (2.5, -11.5) {$\leftarrow$};
\node[align=left] at (0.5,-8.5) {$\uparrow$}; 
\node[align=left] at (1.4,-8.5) {$\leftarrow$};
\node[align=left] at (1.5,-8.4) {$\uparrow$};
\node[align=left] at (3.5,-8.5) {$\uparrow$};
\node[align=left] at (4.4,-8.5) {$\leftarrow$};
\node[align=left] at (4.5,-8.4) {$\uparrow$};
\node[align=left] at (0.5,-10.5) {$\uparrow$};
\node[align=left] at (1.4,-10.5) {$\leftarrow$};
\node[align=left] at (1.5,-10.4) {$\uparrow$};
\node[align=left] at (1.4,-11.5) {$\leftarrow$};
\node[align=left] at (1.5,-11.4) {$\uparrow$};
\node[align=left] at (3.5,-10.5) {$\uparrow$};
\node[align=left] at (3.5,-11.5) {$\uparrow$};
\node[align=left] at (4.4,-10.5) {$\leftarrow$};
\node[align=left] at (4.5,-10.4) {$\uparrow$};
\node[align=left] at (4.4,-11.5) {$\leftarrow$};
\node[align=left] at (4.5,-11.4) {$\uparrow$};
\node[align=left] at (1.5,-7.5) {$\leftarrow$};
\node[align=left] at (2.5,-7.5) {$\leftarrow$};
\node[align=left] at (3.5,-7.5) {$\leftarrow$};
\draw [decorate,decoration={brace,amplitude=5pt}] (-0.5,-5.0) -- (-0.5,-1);
\node[align=left, rotate=90] at (-1,-3) {$n-2$};
\draw [decorate,decoration={brace,amplitude=5pt}] (0,0.5) -- (5,0.5);
\node[align=left] at (2.5,1) {$5$};
\node[align=left] at (6.5,-13) {};
\end{tikzpicture}}
\subfloat[]{
\includegraphics[scale=0.6]{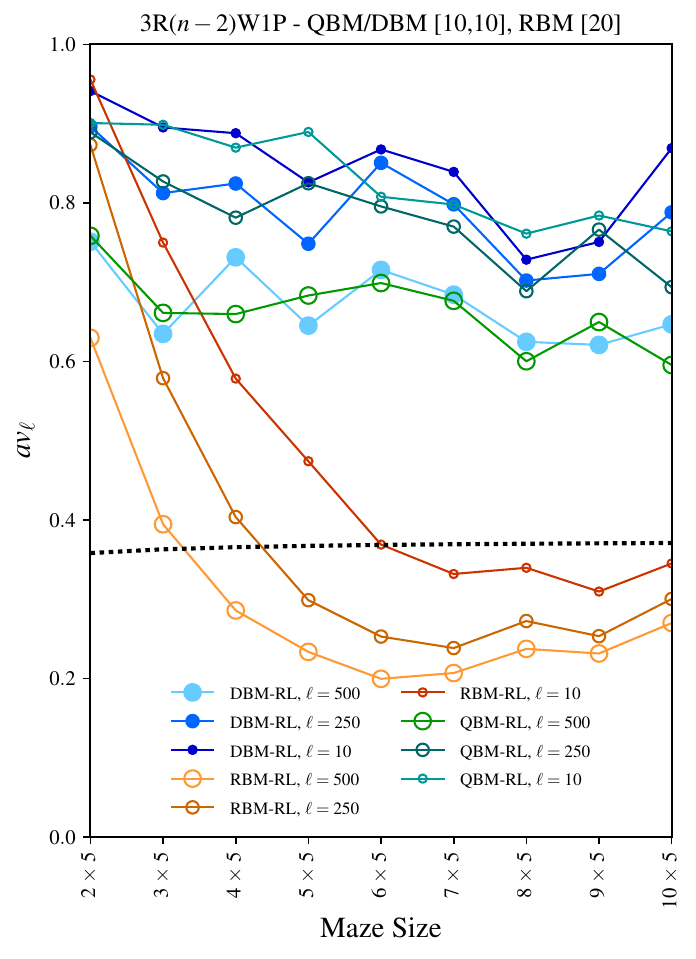}}
\vspace*{13pt}
\captionof{figure}{A comparison between the performance of RBM-RL, DBM-RL, and
QBM-RL as the size of the maze grows. All Boltzmann machines have 20 hidden
nodes. (a) The schematics of an $n \times 5$ maze with one deterministic reward,
2 stochastic rewards, one pit, and $n-2$ walls. (b) The scaling of the average
fidelity of each algorithm run on each instance of the $n \times 5$ maze. The
dotted line is the average fidelity of uniformly randomly generated actions.
\label{nx5}}
\end{figure*}

Our next result, shown in Fig.~\ref{fig:windb}, compares RBM-RL, DBM-RL, and
QBM-RL for a windy maze of size $3 \times 5$. The transition kernel for this
experiment is chosen such that $\mathbb{P}(a(s)|s,a)=0.8,$ and $\mathbb P(s' |
s, a)$ has a nonzero value for all $s' \neq a(s)$ that are reachable from $s$ by
taking some action, in which case all these values are equal. The transition
probability is zero for all other states. Fig.~\ref{fig:wind-ker} shows examples
of the transition probabilities in the windy problem.

To demonstrate the performance of RBM-RL, DBM-RL, and QBM-RL with respect to
scaling, we define another measure called average fidelity, $av_\ell$, where we
take the average fidelity over the last $\ell$ training samples of the fidelity
measure. Given $T_s$ total training samples and $\text{fid}(i)$ as defined
above, we write
\begin{align*}
av_\ell = \frac{1}{\ell} \sum_{i=T_s - \ell}^{T_s} \text{fid}(i)\,.
\end{align*}
In Fig.~\ref{nx5}, we report the effect of maze size on $av_\ell$ for RBM-RL,
DBM-RL, and QBM-RL for varying maze sizes. We plot $av_\ell$ for each algorithm
with $\ell = 500, 250,$ and $10$ as a function of maze size for a family of
problems with one deterministic reward, two stochastic rewards, one pit, and
$n-2$ walls. We use nine $n \times 5$ mazes in this experiment, indexed by
various values of $n$. In addition to the $av_\ell$ plots, we include a
dotted-line
plot depicting the fidelity for a completely random policy. The fidelity of
the random policy is given by the average probability of choosing an optimal
action at each state when generating admissible actions uniformly at random,
which is given by $\frac{18n+7}{48n + 24}$. Note that the fidelity of the
random policy increases as the maze size increases. This is due to the fact that
maze rows containing a wall have more average admissible optimal actions than
the top and bottom rows of the maze.

\section{Discussion}

The fidelity curves in Fig.~\ref{fig:3by5} show that \mbox{DBM-RL} outperforms
\mbox{RBM-RL} with respect to the number of training samples. Therefore, we
expect that in conjunction with a high-performance sampler of Boltzmann
distributions (e.g., a quantum or a quantum-inspired oracle taken as such),
\mbox{DBM-RL} improves the performance of reinforcement learning. \mbox{QBM-RL}
is not only on par with \mbox{DBM-RL}, but actually slightly improves upon it by
taking advantage of sampling in the presence of a significant transverse field.

This is a positive result for the potential of sampling from a quantum device in
machine learning, as we do not expect quantum annealing to obtain the Boltzmann
distribution of a classical Hamiltonian \cite{amin2015searching,
1367-2630-11-7-073021, PhysRevA.95.042302}. However, given the discussion in
Sec.~\ref{sec:open}, a quantum annealer viewed as an open system coupled to a
heat bath could be a better choice of sampler from its instantaneous Hamiltonian
in earlier stages of the annealing process, compared to a sampler of the
problem Hamiltonian at the end of the evolution. Therefore, these experiments
address whether a quantum Boltzmann machine with a transverse field Ising
Hamiltonian can perform at least as well as a classical Boltzmann machine.

In each experiment, the fidelity curves from \mbox{DBM-RL} produced using SQA
with $\Gamma_f = 0.01$ match the ones produced using SA. This is consistent
with our expectation that using SQA with $\Gamma \to 0$ produces samples from
the same distribution as SA, namely, the Boltzmann distribution of the classical
Ising Hamiltonian with no transverse field.

The best algorithm in our experiments is evidently QBM-RL using SQA. Here, the
final transverse field is $\Gamma_f = 2.00$, corresponding to one-third of the
anneal for a quantum annealing algorithm that evolves along the convex linear
combination of the initial and final Hamiltonians with constant speed. This is
consistent with ideas found in \cite{1601.02036} on sampling at 
\textit{freeze-out} \cite{amin2015searching}.

Fig.~\ref{smooth:1R1W1P} shows that, whereas the maze can be solved with fewer
training samples using ordered sweeps of the maze, the periodic behaviour of the
fidelity curves is due to this periodic choice of training samples. This effect
disappears once the training samples are chosen uniformly randomly.

Fig.~\ref{fig:windb} shows that the improvement in the learning of the DBM-RL
and QBM-RL algorithms persists in the case of more-complicated transition
kernels. The same ordering of fidelity curves discussed earlier is observed:
\mbox{QBM-RL} outperforms DBM-RL, and DBM-RL outperforms RBM-RL.

It is worth mentioning that, even though it may seem that more connectivity
between the hidden nodes may allow a Boltzmann machine to capture more-
complicated correlations between the visible nodes, the training process of the
Boltzmann machine becomes more computationally involved. In our reinforcement
learning application, an RBM with $m$ hidden nodes, and $n= |S| + |A|$ visible
nodes, has $mn$ weights to train. A DBM with two hidden layers of equal size has
$\frac{1}4 m( 2 n + m)$ weights to train. Therefore, when $m < 2 n$, the
training of the DBM is in a domain of a lower dimension. Further, a GBM with
all of its hidden nodes forming a complete graph requires $mn + \binom{m}2$
weights to train, which is always larger than that of an RBM or a DBM with the
same number of hidden nodes.

One can observe from Fig.~\ref{nx5} that, as the maze size increases and the
complexity of the reinforcement learning task increases, $av_\ell$ decreases for
each algorithm. The RBM algorithm, while always outperformed by \mbox{DBM-RL}
and \mbox{QBM-RL}, shows a much faster decay in average fidelity as a function
of maze size compared to both DBM-RL and QBM-RL. For larger mazes, the
\mbox{RBM} algorithm fails to capture maze traversal knowledge, and approaches
$av_\ell$ of a random action allocation (the dotted line), whereas the DBM-RL
and QBM-RL algorithms continue to be trained well. DBM-RL and QBM-RL are capable
of training the agent to traverse larger mazes, whereas the RBM algorithm,
utilizing the same number of hidden nodes and a larger number of weights, fails
to converge to an output that is better than a random policy.

The runtime and computational resources needed to compare DBM-RL and QBM-RL with
RBM-RL have not been investigated here. We expect that in view of
\cite{le2008representational}, the size of RBM needed to solve larger maze
problems will grow exponentially. Thus, it would be interesting to research the
extrapolation of the asymptotic complexity and size of the DBM-RL and QBM-RL
algorithms with the aim of attaining a quantum advantage. Applying the
algorithms described in this paper to tasks that have larger state and action
spaces, as well as to more-complicated environments, will allow us to
demonstrate the scalability and usefulness of the DBM-RL and QBM-RL approaches.
The experimental results shown in Fig.~\ref{nx5} represent only a rudimentary
attempt to investigate this matter, yet the results are promising. However, this
experiment does not provide a practical characterization of the scaling of our
approach, and further investigation is needed.

\section*{Acknowledgements}
We would like to thank Hamed~Karimi, Helmut~Katzgraber, Murray~Thom,
Matthias~Troyer, and Ehsan~Zahedinejad, as well as the referees and editorial
board of Quantum Information and Computation, for reviewing this work and
providing many helpful suggestions. The idea of using SQA to run experiments
involving measurements with a nonzero transverse field was communicated in
person by Mohammad Amin. We would also like to thank Marko~Bucyk for editing
this manuscript.

\nocite{*} 

\bibliographystyle{ieeetr}

\end{document}